\documentstyle[fleqn,12pt]{article} 
\begin{document} 
\baselineskip24pt 
\begin{center} 
{\Large\bf Chiral Perturbation Theory and \\ 
$U(3)_L\times U(3)_R$ Chiral Theory of Mesons}\\[5mm] 
Xiao-Jun Wang \\ 
{\small 
Center for Fundamental Physics, 
University of Science and Technology of China\\
Hefei, Anhui 230026, P.R.China\\
and}\\

Mu-Lin Yan\\
{\small
International Centre for Theoretical Physics\\
P.O.Box 586, 34100, Trieste, Italy\\
and\\
Center for Fundamental Physics\\
University of Science and Technology of China\\
Hefei, Anhui 230026\footnote{$^)$Mailing
address, E-mail: mlyan@nsc.ustc.edu.cn}}$^)$\\
(published in J. Phys. G: Nucl. Part. Phys. {\bf 24} (1998) 1077-1088.)\\
\end{center}
\begin{abstract}
\noindent
{\small  We examine low energy limit of
 $U(3)_L\times U(3)_R$ chiral theory of mesons  
 through integrating out fields of vector and axial-vector mesons. 
 The effective lagrangian for pseudoscalar mesons 
 at $O(p^4)$ has been obtained, and five
 low energy coupling  constants $L_i(i=1,2,3,9,10)$
 have been revealed. They are in good 
 agreement with the results of  $\chi$PT's at $\mu \sim m_\rho$. }
\end{abstract}

   {{\bf 1. Introduction}}
   
   It is well known that, at low energies
(e.g.,the energy scale is $\mu \sim m_\rho$), the strong,
electromagnetic
and weak interactions of pseudoscalar mesons can be successfully described
by the chiral perturbation theory($\chi$PT)\cite{Daschen, Gasser}. 
This effective
theory depends on a number of low-energy coupling constants which cannot
be determined from the symmetries of the fundamental theory only. They are
in principle determined by the underlying QCD dynamics in terms of
renormalization group invariant scale $\Lambda_{QCD}$ and the heavy quark masses.
The foundations of $\chi$PT has been discussed in ref.\cite{Leutwyler}.
Actually, $\chi$PT is the effective theory of QCD at low-energies, and any
eligible low-energy theories of mesons have to be consistent with it. By
using this claim as a criterion, 
the authors of ref.\cite{Ecker, Donoghue} have examined,
commented on or criticized several chiral theoretical models of mesons. In
particular, it was pointed in ref.\cite{Ecker, Ecker1} that a reliable
low energy effective theory which includes meson resonances must return 
to $\chi$PT when the freedom of meson resonance is frozen, or 
``integrated out''. The remainder effects of these meson resonances 
(or effects of virtual mesons) are absorbed by the low-energy coupling 
constances (or chiral coupling constances) of $\chi$PT. 
In other words, the low energy limit of effective meson theories with
excited meson resonances has to be consistent with $\chi$PT.

Recently, based on $U(3)_L \times U(3)_R$ chiral symmetry and an ansatz
that mesons are composite fields of quarks, Li\cite{BAL1}-\cite{Gao}  have 
proposed an effective meson theory with spin 1 meson resonances
(hereafter we shall call this theory as the Li model).
Distinguishing from other chiral theories, a cutoff $\Lambda \sim 1.6GeV$
\footnote{$^)$See eqs.(129) and (130) in ref.\cite{BAL1}. 
In the present paper, a universal
constant g relate to the cut-off $\Lambda$ through
$$
g^2=\frac{8}{3}
\frac{N_c}{(4\pi)^2}\frac{D}{4} \Gamma(2-\frac{D}{2})(\frac{\mu^2}{m^2})
^{\frac{\epsilon}{2}}
$$ 
This constant $g$ can be determinded
from decay $\rho \rightarrow \pi\pi$. 
As $g=0.35$, the cut-off $\Lambda \sim 1.6GeV$ .}$^)$
(which is higher than $m_\rho =0.77GeV$) has been
introduced as an intrinsic parameter of this truncated field theory of 
mesons, and all the masses of the mesons in the model are 
below this cutoff $\Lambda$. This fact indicates that Li model is 
a self-consistent meson theory at intermedate energy scale. In this present 
paper we try to study the low energy limit of Li model and to see whether
it is consistent with $\chi$PT or not. Specifically, we will calculate 
the chiral coupling constants $L_1$, $L_2$, $L_3$, $L_9$, $L_{10}$ and
$H_1$ at $\mu \sim m_\rho$ by using the path
integration to freeze the field freedoms of vector and axial-vector
meson resonances (or "integrate out" those meson fields) in the Li model.
Because $L_1, L_2,
L_3, L_9$ and $L_{10}$ are related to five conditions abstracted from 
QCD\cite{Ecker1}, 
our calculations in this paper could be thought of as a test of the Li model
from the QCD points of view. Therefore, even though the phenomenological predictions
have been extensively studied and met with remarkable 
success\cite{BAL1,BAL2,BAL3,
BAL4,Gao}, the examination to the model's low energy limit is 
still necessary.

 In ref.\cite{Ecker1}, the effects of virtual meson resonances in several
models have been calculated. In those calculations
only the terms with one meson resonance
in interactions were taken into account. Under this approximation, it is
enough for the purpose there to construct the meson resonance fields by
simply using the propagators.
The corresponding virtual meson resonance effects are due to the exchanges 
of these virtual particles.
However, this method fails to the models in which the terms
with higher power of resonances have contributions to the interactions 
at $O(p^4)$
\footnote{$^)$
The discussions
in ref.\cite{Ecker1} on the models involving massive Yang-Mills fields
(see Meissner in \cite{MB}) or hidden-gauge
(see Bando et.al. in \cite{MB}) are incomplete,
because the terms with higher powers of $V$ at $O(p^4)$,  
such as $VVV$, $VVVV$ are ignored.}$^)$.
In Li model, the vector($V$) and the axial vector($A$) mesons are ordinary
4-component vectors. The terms with more than one resonance fields, 
such as $VV$, $VVV$, $VVVV$, $VVAA$, $AA$, $AAAA$, $AAA\partial \phi$ 
( $\phi$ is pseudoscalar meson) and so on, will emerge in the model's 
interactions at $O(p^4)$. Thus, in order to catch at the effects 
at $O(p^4)$ of all virtual meson
resonances in Li model, we have to carry out the corresponding non-Gaussian
type path integrations to $V$ and $A$ under appropriate approximations.
Practically, we will use the perturbation expansions around 
the saddle point of the model's Lagrangian to perform 
the path integrations for our purpose. 
In this manner, the effects of virtual meson resonances are divided into two
parts: effects of virtual resonance exchanges (EVRE) and effects of virtual
resonance vertices (EVRV).
Actually, the corresponding calculations in ref.\cite{Ecker1} are
equivalent to performing Gaussian type path integrations, and 
one of the things done
in this paper is to extend the ref.\cite{Ecker1} to the non-Gaussian
type case. The former evaluates EVRE only, and the latter evaluates both.
In addition, we will also show (see section 3)
the effects of quark constitutes of pseudoscalar mesons (EQCPM)
in Li model contributes to $L_i (i=1, 2, 3, 9, 10)$ at $O(p^4)$
besides EVRE and EVRV.
All EVRE, EVRV and EQCPM will be taken into consideration in the calculations
of $L_i$ presented in this paper.

In sections 2 and 3, the notations and definitions in $\chi$PT and Li model
are presented respectively. The section 4 devotes to calculate the low energy
coupling constants. Finally, we discuss the results of this paper 
in section 5.

\vspace{8mm}
 {{\bf 2. Chiral Perturbation Theory}}
 
 The effective lagrangian of chiral perturbation theory
can be found in ref.\cite{Gasser}
\begin{equation}
\label{e1}
        {\cal L}_{\chi PT}={\cal L}_2+{\cal L}_4+...
\end{equation}
${\cal L}_2$ is the nonlinear $\sigma$ model lagrangian coupled to external
fields $\upsilon,a,s$ and $p$
\begin{equation}
\label{e2}
       {\cal L}_2=\frac{f_{\pi} ^2}{16}<\nabla_\mu U{\nabla^\mu} U^{\dag}+
       {\chi} U^{\dag}+\chi^{\dag} U>,
\end{equation}
where
   \begin{eqnarray}
   \label{e3}
       &&\nabla_\mu U=\partial_\mu U-i(\upsilon_\mu+a_{\mu})U+iU(\upsilon_\mu-
                    a_{\mu}); \\
       &&\chi=2B_0(s+ip)\;\; {\sim}\;\; 2m_qB_0, \nonumber
   \end{eqnarray}
and $<M>$ stands for the trace of the matrix $M$. The external fields
$\upsilon_{\mu},a_{\mu}, s$ and $p$ are Hermitian $3\times 3$ matrices
in flavour space and are vector, axial-vector, scalar and pseudoscalar
external fields respectively. $U$ is a unitary $3\times 3$ matrix
\begin{equation}
\label{e4}
  U=exp(-i\frac{2\Phi}{f_\pi}),\;\;\;\;\;\;\;\;  \Phi=\sum\limits_{i=1}^{8}
     \lambda_{i}\phi^{i}.
\end{equation}
where $f_\pi=186 MeV$.
   The lagrangian (1) exhibits a local $SU(3)_L \otimes SU(3)_R$ symmetry
$$   U {\rightarrow} g_L Ug_R^{\dag}, $$
$$   \upsilon_\mu\pm a_{\mu} {\rightarrow}g_{L.R}(\upsilon_\mu\pm a_{\mu})
      g_{L.R}^{\dag}+ig_{L.R}\partial_\mu g_{L.R}^{\dag},$$
$$   s+ip \rightarrow g_L(s+ip)g_R ^{\dag},              $$
where $g_L,g_R \in SU(3)_L \otimes SU(3)_R$.
The general lagrangian ${\cal L}_4$ is to be used only at tree-level
in $\chi$PT, and all the states to which it is applied obey the equation of
motion
\begin{equation}
\label{e5}
  \nabla_\mu(U{\nabla^{\mu}}U^{\dag})+\frac{1}{2}(\chi^{\dag}U-U^{\dag}{\chi})=0.
\end{equation}
At chiral limit $\chi {\rightarrow}0$, eq.(~\ref{e5})
becomes
\begin{equation}
\label{e6}
   \nabla_\mu U{\nabla^{\mu}} U^{\dag}+U{\nabla_\mu\nabla^{\mu}} U^{\dag}=0.
\end{equation}

In this paper we are concerned with the effective lagrangian at order $p^4$,
\begin{eqnarray}
\label{e7}
  {\cal L}_4&=&L_1<\nabla_{\mu}U\nabla^{\mu}U^{\dag}>^2+L_2<\nabla_{\mu}U\nabla_{\nu}U^{\dag}>
             <\nabla^{\mu}U\nabla^{\nu}U^{\dag}> \nonumber \\
            &\;&+L_3<\nabla_{\mu}U\nabla^{\mu}U^{\dag}\nabla_{\nu}U\nabla^{\nu}U^{\dag}>
             +L_4<\nabla_{\mu}U\nabla^{\mu}U^{\dag}><\chi^{\dag} U+\chi U^{\dag}>\nonumber \\
            &\;&+L_5<\nabla_{\mu}U\nabla^{\mu}(\chi^{\dag}U+\chi U^{\dag})>+
             L_6<\chi^{\dag}U+\chi U^{\dag}>^2 \nonumber \\
            &\;&+L_7<\chi^{\dag}U-\chi U^{\dag}>^2
            +L_8<\chi^{\dag}U{\chi^{\dag}}U+\chi U^{\dag}\chi U^{\dag}>
            \nonumber \\
            &\:&-iL_9<F^{\mu\nu}_{R}\nabla_{\mu}U^{\dag}\nabla_{\nu}U+
             F^{\mu\nu}_{L}\nabla_{\mu}U\nabla_{\nu}U^{\dag}>
             +L_{10}<UF^{\mu\nu}_{R}U^{\dag}F_{L\mu\nu}> \nonumber \\
             &\;&+H_1<F_{R\mu\nu}F^{\mu\nu}_{R}+F_{L\mu\nu}F^{\mu\nu}_{L}>
             +H_2<\chi^{\dag}\chi>,
\end{eqnarray}
where
\begin{equation}
\label{e8}
   F^{\mu\nu}_{L.R}=\partial^{\mu}(\upsilon^{\nu}\mp a^{\nu})
        -\partial^{\nu}(\upsilon^{\mu}\mp a^{\mu})
        -i[\upsilon^{\mu}\mp a^{\mu},\upsilon^{\nu}\mp a^{\nu}].
\end{equation}
$L_1,...,L_{10}$ are ten real low-energy coupling constants at
order $p^4$ which
together with $f_\pi$ and $B_0$ determined completely the low-energy behavior
of pseudoscalar meson interaction to $O(p^4)$.
It is easy to see that $L_1, L_2, L_3, L_9, L_{10}$ and $H_1$ are
independent of quark-masses.  Therefore they could
receive contributions  from the chiral
symmetric models beyond $\chi$PT.

\vspace{8mm}

  {\bf 3. $U(3)_L \times U(3)_R$ Chiral Theory of Mesons }
  
  The lagrangian of Li model with 3-flavor reads\cite{BAL2}
\begin{eqnarray}
\label{e9}
{\cal L}&=&\bar{\psi}(x)(i\gamma\cdot\partial+\gamma\cdot {\cal V}
           +e_0Q\gamma\cdot A+\gamma\cdot {\cal A}\gamma_5-mu(x))\psi(x)
           \nonumber \\
           &\;&+\frac{1}{2}m_0^2({\cal V}_{\mu}^i{\cal V}^{i\mu}+{\cal A}_{\mu}^i
           {\cal A}^{i\mu})
         +\bar{\psi}(x)_L\gamma\cdot W\psi(x)_L,
\end{eqnarray}
where $\psi$ are quark fields $u, d$ and $s$,
${\cal V}$ and ${\cal A}$ are vector and  axial-vector mesons fields respectively, 
$A$ is the photon field, $Q$ is the electric charge operator of $u,d,s$ quarks,
$W_{\mu}^i$ is
$W$ boson. More explicitly, 
 ${\cal V},{\cal A},$ and $u$ can be written as follows\cite{BAL2}
\begin{eqnarray*}
{\cal A}_{\mu}&=&\lambda_i {\cal A}_{\mu}^i=\tau_n a^n_{\mu}+\lambda_a
     K^a_{1\mu}+(\frac{2}{3}+\frac{1}{\sqrt{3}}\lambda_8)f_{\mu}+
          (\frac{2}{3}-\frac{1}{\sqrt{3}}\lambda_8)f_{s\mu},\\
{\cal V}_{\mu}&=&\lambda_i {\cal V}_{\mu}^i=\tau_n \rho^n_{\mu}+\lambda_a
    K^{\ast a}_{\mu}+(\frac{2}{3}+\frac{1}{\sqrt{3}}\lambda_8)\omega_{\mu}+
          (\frac{2}{3}-\frac{1}{\sqrt{3}}\lambda_8)\sqrt{2}\phi_{\mu},\\
u&=&\frac{1}{2}(1+\gamma_5)U+\frac{1}{2}(1-\gamma_5)U^{\dag},
\end{eqnarray*}
\begin{eqnarray*}
 m_0^2({\cal V}_{\mu}^i{\cal V}^{i\mu}+{\cal A}_{\mu}^i{\cal A}^{i\mu})&
      =&m_1^2(\rho_i^{\mu}\rho_{i\mu}+
        \omega^{\mu}\omega_{\mu}+a_i^{\mu} a_{i\mu}+f^{\mu}f_{\mu})
        +m_2^2(K_{\mu}^{\ast a}K^{\ast a\mu}+K_1^{\mu}K_{1\mu}) \\
       &\;&+m_3^2(\phi_{\mu}\phi^{\mu}+f_s^{\mu}f_{s\mu}),
\end{eqnarray*}
where $i=1,...8$; $n=1,2,3$; $a=4,5,6,7$ and $U$ is the same as in eq.(~\ref{e4}). The 
$U_A(1)$ problem is not taken into consideration in this paper.
All meson fields ${\cal V}$, ${\cal A}$ and $u(x)$ emerge in the Lagrangian 
of eq.(9) as auxiliary fields, because there are no kinetic terms to
them. By solving the equations of motion of the fields, one can find
that these auxiliary fields are composed fields of quark
fields\cite{BAL2}. This reflects the basic physics fact that all mesons
are the bound states of quarks.

  By means of path integral, the quark fields can be integrated out, the
effective lagrangian of meson is obtained
$$ {\cal L}(U,{\cal V},{\cal A})={\cal L}_{Re}+{\cal L}_{Im}$$
The lagrangian $L_{Im}$ describes the physical processes with abnormal
parity which will not be discussed in this paper. To the fourth order in
the covariant derivatives in Minkowski space, the lagrangian 
${\cal L}_{Re}$ which
describes the physical processes with normal parity takes the form
\begin{eqnarray}
\label{e10}
  {\cal L}_{Re}&=&\frac{F^2}{16}<D_{\mu}UD^{\mu}U^{\dag}>-\frac{g^2}{8}
                <{\cal V}_{\mu\nu}{\cal V}^{\mu\nu}+{\cal A}_{\mu\nu}
                {\cal A}^{\mu\nu}> \nonumber \\
               &\;&-\frac{3i}{2}\gamma<(D_{\mu}UD_{\nu}U^{\dag}
                +D_{\mu}U^{\dag}D_{\nu}U){\cal V}^{\mu\nu}> \nonumber \\
               &\;&-\frac{3i}{2}\gamma<(D_{\mu}U^{\dag}D_{\nu}U
                -D_{\mu}UD_{\nu}U^{\dag}){\cal A}^{\mu\nu}> \nonumber \\
               &\;&+\frac{\gamma}{2}<D_{\mu}D_{\nu}UD^{\mu}D^{\nu}U^{\dag}>
                +\frac{1}{4}m_0^2<{\cal V}_{\mu}{\cal V}^{\mu}
                +{\cal A}_{\mu}{\cal A}^{\mu}> \nonumber \\
               &\;&+\frac{\gamma}{4}<D_{\mu}UD_{\nu}U^{\dag}D^{\mu}U
                D^{\nu}U^{\dag}-2D_{\mu}UD^{\mu}U^{\dag}D_{\nu}U
                D^{\nu}U^{\dag}>
\end{eqnarray}
with $$\gamma=\frac{N_c}{3(4\pi)^2}$$ and
\begin{eqnarray}
\label{e11}
&&D_{\mu}U=\nabla_{\mu}U-i({\cal V}_{\mu}-{\cal A}_{\mu})U
    +iU({\cal V}_{\mu}+{\cal A}_{\mu}), \nonumber \\
&&{\cal V}_{\mu\nu}=\partial_{\mu}({\cal V}_{\nu}+\upsilon_{\nu})
     -\partial_{\nu}({\cal V}_{\mu}+\upsilon_{\mu})-i[{\cal V}_{\mu}
     +\upsilon_{\mu},{\cal V}_{\nu}+\upsilon_{\nu}]
     -i[{\cal A}_{\mu}+a_{\mu},{\cal A}_{\nu}+a_{\nu}],  \\
&&{\cal A}_{\mu\nu}=\partial_{\mu}({\cal A}_{\nu}+a_{\nu})-\partial_{\nu}
     ({\cal A}_{\mu}+a_{\mu})-i[{\cal A}_{\mu}+a_{\mu},{\cal V}_{\nu}+\upsilon_{\nu}]
     -i[{\cal V}_{\mu}+\upsilon_{\mu},{\cal A}_{\nu}+a_{\nu}], \nonumber \\
&&D_{\mu}D_{\nu}U=\nabla_{\mu}(\nabla_{\nu}U)-i({\cal V}_{\mu}-{\cal A}_{\mu})D_{\nu}U
    +iD_{\nu}U({\cal V}_{\mu}+{\cal A}_{\mu}), \nonumber
\end{eqnarray}
in this form, $\nabla_{\mu}U$ is the same as eq.(~\ref{e3}), there are only
three independent parameters in this theory and they are 
$g,f_\pi,m_\rho$ ($m_0=g m_\rho$). It is should be noted $F \neq f_{\pi}$ in
lagrangian (10) because of mixing between field ${\cal A}_{\mu}(x)$ and
$\partial_{\mu} \pi(x)$, which should be diagonalized via field redefinition as,
$$ {\cal A}_{\mu}(x) \rightarrow {\cal A}_{\mu}(x)+c \partial_{\mu} \pi(x)$$
the following equations has been given in Ref.\cite{BAL1},
$$\frac{F^2}{f_\pi^2}(1-\frac{c}{g})=1$$
$$c=\frac{f_\pi^2}{2gm_\rho^2}$$.

Eq.(10) is the meson Lagrangian of Li model at energy scale $\Lambda
\sim 1.6GeV$. Besides the pseudoscalar mesons, the vector ($V$)
and axial vector ($A$) meson resonances are excited in eq.(10). Our object is to reduce
it to an effective Lagrangian containing pseudoscalar mesons only by 
integrating out $V$ and $A$, to reveal the values of
$L_1, L_2, L_3, L_9, L_{10}$ and $H_1$. The interactions of $V$ and
$A$ in eq.(10) will contribute to
these low energy coupling constants. Because the
terms with higher powers of $V$ and $A$,
such as $VV$, $VVV$, $VVVV$, $VVAA$, $AA$, $AAAA$, $AAA\partial \phi$ 
( $\phi$ is pseudoscalar meson) and so on, emerge in the interactions
of eq.(10). As stated in section 1, we
have to calculate both EVRE and EVRV. In addition, in eq.(10), there are 
terms with 4 derivatives to pseudoscalar fields and independent of
$V$ and $A$. The coupling coefficients of these terms contribute to
$L_i, (i=1,2,3,9,10)$ also. They reflect the direct contributions due to
$L_i$ of the effects of quark constitutes of pseudoscalar mesons (EQCPM)
in the Li model. Consequently, the low-energy coupling constants at $O(p^4)$
for Li model receive contributions from three effects: EVRE, EVRV and EQCPM.
Hereafter, we will take all effects into consideration.

\vspace{8mm}

 {{\bf 4. Calculations of Low Energy Coupling Constants}} 

In this section, we will use path integration manner to integrate out
field freedoms of $V$ and $A$ in ${\cal L}_{Re}$(~\ref{e10}),
and to derive the chiral effective Lagrangian for pseudoscalar
mesons, which is the low energy limit of the Li model. 
Comparing it with the standard $\chi$PT 
Lagrangian of eq.(7), we then work out
the low-energy coupling constants at order $p^4$ for Li model.

For the sake of convenience we define
\begin{eqnarray*}
   &&L_{\mu}=\frac{1}{\sqrt{2}}({\cal V}_{\mu}-{\cal A}_{\mu}), \;\;\;\;\;
    R_{\mu}=\frac{1}{\sqrt{2}}({\cal V}_{\mu}+{\cal A}_{\mu}), \\
   &&l_{\mu}=\frac{1}{\sqrt{2}}(\upsilon_{\mu}-a_{\mu}), \;\;\;\;\;
    r_{\mu}=\frac{1}{\sqrt{2}}(\upsilon_{\mu}+a_{\mu}), \\
   &&L_{\mu\nu}=\frac{1}{\sqrt{2}}({\cal V}_{\mu\nu}-{\cal A}_{\mu\nu})=\nabla_{\mu}^{L}L_{\nu}
              -\nabla_{\nu}^{L}L_{\mu}-\sqrt{2}i[L_{\mu},L_{\nu}], \\
   &&R_{\mu\nu}=\frac{1}{\sqrt{2}}({\cal V}_{\mu\nu}+{\cal A}_{\mu\nu})=\nabla_{\mu}^{R}R_{\nu}
              -\nabla_{\nu}^{R}R_{\mu}-\sqrt{2}i[R_{\mu},R_{\nu}],
\end{eqnarray*}
where
\begin{eqnarray*}
   &&\nabla_{\mu}^{L}L_{\nu}=\partial_{\mu}L_{\nu}-\frac{1}{\sqrt{2}}i[l_{\mu},
         L_{\nu}],  \\
   &&\nabla_{\mu}^{R}R_{\nu}=\partial_{\mu}R_{\nu}-\frac{1}{\sqrt{2}}i[r_{\mu},
         R_{\nu}].
\end{eqnarray*}
Then we have
\begin{eqnarray}
\label{e12}
   &&F_{\mu\nu}^{L}=\sqrt{2}(\partial_{\mu}l_{\nu}-\partial_{\nu}l_{\mu})
       -2i[l_{\mu},l_{\nu}], \nonumber \\
   &&F_{\mu\nu}^{R}=\sqrt{2}(\partial_{\mu}r_{\nu}-\partial_{\nu}r_{\mu})
       -2i[r_{\mu},r_{\nu}],
\end{eqnarray}
and
\begin{eqnarray}
\label{e13}
    &&[\nabla_{\mu},\nabla_{\nu}]U=i(UF_{\mu\nu}^R-F_{\mu\nu}^LU),
      \nonumber \\
    &&[\nabla_{\mu},\nabla_{\nu}]U^{\dag}=i(U^{\dag}F_{\mu\nu}^L
          -F_{\mu\nu}^R U^{\dag}).
\end{eqnarray}
With the above notations, the lagrangian (~\ref{e10}) can be written as follows
\begin{eqnarray}
\label{e14}
  {\cal L}_{Re}&=&\frac{F^2}{16}<D_{\mu}UD^{\mu}U^{\dag}>-\frac{g^2}{16}
                <2L_{\mu\nu}L^{\mu\nu}+2R_{\mu\nu}R^{\mu\nu}+F_{\mu\nu}^{L}
                F^{L\mu\nu}+F_{\mu\nu}^{R}F^{R\mu\nu}> \nonumber \\
               &\;&-\frac{g^2}{4\sqrt{2}}<L_{\mu\nu}F^{L\mu\nu}
                +R_{\mu\nu}F^{R\mu\nu}> \nonumber \\
                &\;&-\frac{3i}{\sqrt{2}}\gamma<D_{\mu}UD_{\nu}U^{\dag}
                L^{\mu\nu}+D_{\mu}U^{\dag}D_{\nu}UR^{\mu\nu}> \nonumber \\
                &\;&-\frac{3i}{2}\gamma<D_{\mu}U^{\dag}D_{\nu}U
                F^{L\mu\nu}+D_{\mu}UD_{\nu}U^{\dag}F^{R\mu\nu}> \nonumber \\
                &\;&+\frac{\gamma}{2}<D_{\mu}D_{\nu}UD^{\mu}D^{\nu}U^{\dag}>
                +\frac{1}{4}m_0^2<L_{\mu}L^{\mu}+R_{\mu}R^{\mu}> \nonumber \\
                &\;&+\frac{\gamma}{4}<D_{\mu}UD_{\nu}U^{\dag}D^{\mu}U
                D^{\nu}U^{\dag}-2D_{\mu}UD^{\mu}U^{\dag}D_{\nu}U
                D^{\nu}U^{\dag}>,
\end{eqnarray}
where
\begin{eqnarray*}
  &&\nabla_{\mu}U=\partial_{\mu}U-\sqrt{2}il_{\mu}U+\sqrt{2}iUr_{\mu}, \\
  &&D_{\mu}U=\nabla_{\mu}U-\sqrt{2}iL_{\mu}U+\sqrt{2}iUR_{\mu}.
\end{eqnarray*}

All vector and axial-vector meson resonances in Li model join meson
dynamics via $L_{\mu}$, $R_{\mu}$ in ${\cal L}_{Re}$ (eq.(~\ref{e14})).
To order $p^4$ in chiral expansion, the virtual particle effects of 
these spin-1 mesons will
induce a local lagrangian of type eq.(~\ref{e7}) with their contributions to
$L_i$. Our objective is to derive this lagrangian by completing path
integration over $L_{\mu}$ and $R_{\mu}$ in ${\cal L}_{Re}$. For this
purpose, we divide ${\cal L}_{Re}$ into two parts,
$$ {\cal L}_{Re}=({\cal L}_{Re})_2+({\cal L}_{Re})_{3,4}, $$
where $({\cal L}_{Re})_2 $ contains the terms up to the second
power of $L_{\mu}$ and $R_\mu$, and $({\cal L}_{Re})_{3,4}$, 
the terms with third and fourth power of them. 
$({\cal L}_{Re})_2$ can be rewritten as
\begin{eqnarray}
\label{e15}
   ({\cal L}_{Re})_2&=&{\cal L}^{(0)}(U)+<J_{\mu}^{L}L^{\mu}>
     +<J_{\mu}^{R}R^{\mu}>+<S_{\mu}^{L}(U,R)L^{\mu}> \nonumber \\
     &\;&-\frac{1}{2}<L_{\mu}M_{L}^{\mu\nu}L_{\nu}+R_{\mu}M_{R}^{\mu\nu}R_{\nu}>,
\end{eqnarray}
where
\begin{eqnarray}
\label{e16}
  &&{\cal L}^{(0)}(U)=\frac{F^2}{16}<\nabla_{\mu}U\nabla^{\mu}U^{\dag}>,
  \nonumber \\
  &&J_{\mu}^L=-\frac{\sqrt{2}}{8}F^2iU\nabla_{\mu}U^{\dag}, \nonumber \\
  &&J_{\mu}^R=-\frac{\sqrt{2}}{8}F^2iU^{\dag}\nabla_{\mu}U,   \\
  &&S_{\mu}^L(U,R)=-\frac{F^2}{4}UR_{\mu}U^{\dag}+2\gamma\nabla_{\nu}^L
        (U\nabla_R^{\nu}R_{\mu}U^{\dag}), \nonumber \\
  &&M^{\mu\nu}_{L.R.}=-(\frac{g^2}{2}-2\gamma)\nabla^2_{L.R.}g^{\mu\nu}
         -(\frac{F^2}{4}+\frac{m_0^2}{2})g^{\mu\nu}+\frac{g^2}{2}
         \nabla^{\mu}_{L.R.}\nabla^{\nu}_{L.R.}. \nonumber
\end{eqnarray}
Noting the symmetry of $L_{\mu}$ and $R_{\mu}$ in ${\cal L}_{Re}$, 
we have
  $$<S_{\mu}^L(U,R)L^{\mu}>=<S_{\mu}^R(U.L)R^{\mu}>, $$
where
  $$S_{\mu}^R(U,L)=-\frac{F^2}{4}U^{\dag}L_{\mu}U+2\gamma\nabla_{\nu}^R
       (U^{\dag}\nabla_L^{\nu}L_{\mu}U). $$
The terms with the third and the fourth power are
$$ ({\cal L}_{Re})_{3,4}={\cal L}_{Re}-({\cal L}_{Re})_2 
=({\cal L}_{Re})_{3,4}(U,L,R). $$
We now evaluate both the exchange effects and the vertex effects
of virtual spin-1 meson resonances (or $L$ and $R$)
in ${\cal L}_{Re}$ and work out the desired effective lagrangian 
${\cal L}_{eff}$ which contains pseudoscalar mesons only. 
The functional integral expression for ${\cal L}_{eff}$ is
\begin{eqnarray*}
   &\;&\exp{\{i\int d^{4}x{\cal L}_{eff}(U)\}} \nonumber \\
   &=&\int [dL][dR]\exp{\{i\int d^{4}x{\cal L}_{Re}\}}   \nonumber \\
          &=&\exp{\{i\int d^{4}x({\cal L}_{Re})_{3,4}(U,\frac{\delta}{\delta J_{\mu}^{L}},
          \frac{\delta}{\delta J_{\mu}^{R}})\}}\int [dL][dR]\exp{\{i\int d^4x
          ({\cal L}_{Re})_2\}}.
\end{eqnarray*}
Owing to Gaussian integral formula, the functional integrations over $L$ and
$R$ in above equation can be performed exactly, and $({\cal L}_{Re})_{3,4}$
can be treated perturbatively.

From $({\cal L}_{Re})_2$, the classical field equations of $L$ and $R$
read
\begin{eqnarray}
\label{e17}
    M^{\mu\nu}_{L}L_{\nu}=J^{\mu}_L+S^{\mu}_L(U,R),  \nonumber \\
    M^{\mu\nu}_{R}R_{\nu}=J^{\mu}_R+S^{\mu}_R(U,L).
\end{eqnarray}
Substituting eq.(~\ref{e16}) into eq.(~\ref{e17}), at $O(p^1)$ the solutions of
eq.(~\ref{e17}) are given as
\begin{eqnarray}
\label{e18}
   L_{\mu}^{(1)c}=kJ_{\mu}^{L}=\frac{i}{\sqrt{2}}\frac{\beta}{g}U\nabla_{\mu}
                   U^{\dag},  \nonumber \\
   R_{\mu}^{(1)c}=kR_{\mu}^{R}=\frac{i}{\sqrt{2}}\frac{\beta}{g}U^{\dag}
                   \nabla_{\mu}U.
\end{eqnarray}
where
\begin{equation}
\label{e19}
   k=-\frac{2}{m_0^2+F^2}.
\end{equation}
For comparing with ref.\cite{BAL1}(similar to the quantity $c$ in \cite{BAL1}) 
we define
\begin{equation}
\label{e20}
   \beta=\frac{gF^2}{2(m_0^2+F^2)}.
\end{equation}
Furthermore, the solution of eq.(~\ref{e17}) at $O(p^3)$ are
\begin{eqnarray}
\label{e21}
   &&L_{\mu}^{(3)c}=L_{\mu}^{(1)c}+\alpha\Theta_{\mu},  \nonumber \\
   &&R_{\mu}^{(3)c}=R_{\mu}^{(1)c}+\alpha\Omega_{\mu},
\end{eqnarray}
where
\begin{equation}
\label{e22}
        \alpha=-\frac{2}{m_0^2}(1-\frac{F^2}{m_0^2}),
\end{equation}
and index $c$ denote classical solution of eq.(~\ref{e17})
\begin{eqnarray*}
   \Theta_{\mu}=(\frac{g^2}{2}-2\gamma)\nabla_{L}^2 L_{\mu}^{(1)c}
       -\frac{g^2}{2}\nabla_{\nu}^{L}\nabla_{\mu}^{L} L^{\nu(1)c}
       +2\gamma\nabla_{L}^{\nu}(U\nabla_{\nu}^{R} R_{\mu}^{(1)c}U^{\dag}),
       \nonumber \\
   \Omega_{\mu}=(\frac{g^2}{2}-2\gamma)\nabla_{R}^2 R_{\mu}^{(1)c}
       -\frac{g^2}{2}\nabla_{\nu}^{R}\nabla_{\mu}^{R} R^{\nu(1)c}
       +2\gamma\nabla_{R}^{\nu}(U\nabla_{\nu}^{L} L_{\mu}^{(1)c}U^{\dag}).
\end{eqnarray*}

Thus, with $L^c$ and $R^c $, we obtain the
${\cal L}_{eff}$ to order $p^4$
\begin{eqnarray}
\label{e23}
    {\cal L}_{eff}&=&({\cal L}_{eff})_2+({\cal L}_{eff})_4+O(p^6) \nonumber \\
&=&{\cal L}^{(0)}(U)+\frac{1}{2}<J_{\mu}^{L}L^{\mu(3)c}+
        J_{\mu}^{R}R^{\mu(3)c}> \nonumber \\
       &\;&+\frac{1}{4}<S_{\mu}^{L}(U,R^c)L^{\mu c}>+({\cal L})_{3,4}(U,
        \frac{R_{\mu}^{(1)c}}{2},\frac{L_{\mu}^{(1)c}}{2})+O(p^6).
\end{eqnarray}
Substituting eq.(~\ref{e18})(~\ref{e21}) into the lagrangian(~\ref{e23}), the
contributions of lagrangian ${\cal L}_{Re}$ (eq.(~\ref{e10})) to the coupling
constants at the order $p^2$ can be found from the following expression,
\begin{equation}
\label{e24}
   ({\cal L}_{eff})_2=\frac{F^2}{16}(1-\frac{2\beta}{g}(1-\frac{\beta}{4g}))
          <\nabla_{\mu}U\nabla^{\mu}U^{\dag}>.
\end{equation}
 This is the lagrangian of kinetic energies for pseudoscalar meson after 
integrating out ${\cal V}_\mu$ and ${\cal A}_\mu$ in the
Li model. The first two terms are similar to the original Li
model (before the integration), however the third one
$\frac{F^2}{16}\frac{\beta}{2g^2}<\nabla_{\mu}U\nabla^{\mu}U^{\dag}>$
is new. It comes from the mixed term of left-hand and right-hand fields
coupling $<S^L_{\mu}(U,R)L^{\mu}>$ of $({\cal L}_{Re})_2$(eq.(~\ref{e15})).
In the original Li model\cite{BAL1},
$<S^L_{\mu}(U,R)L^{\mu}>$ contributes to the
three-point vertex of pseudoscalar, vector and axial-vector mesons instead of
a kinetic term. However, when the vector and axial-vector mesonic fields
were integrated out, this interaction contributes to $({\cal L}_{eff})_2$.
In general, the form of the lagrangian should be different,
since the lagrangian of Li model contains the vector and axial vector resonance
mesons dynamical fields which mimic these dynamics in the low energy region.

To normalize the kinetic energy term of pseudoscalar mesons, we have
\begin{equation}
\label{e25}
    \frac{F^2}{f_{\pi}^2}(1-\frac{2\beta}{g}(1-\frac{\beta}{4g}))=1.
\end{equation}
Parameter $F$ can be fixed by eqs.(~\ref{e20}) and (~\ref{e25}) as long as $g$
and $f_{\pi}$ are known.

The following equations can be easyly proved
\begin{eqnarray*}
  &&\nabla_{\mu}^L(U\nabla_{\nu}U^{\dag})=\nabla_{\mu}U\nabla_{\nu}U^{\dag}
             +U\nabla_{\mu}\nabla_{\nu}U^{\dag}, \\
  &&\nabla_{\mu}^R(U^{\dag}\nabla_{\nu}U)=\nabla_{\mu}U^{\dag}\nabla_{\nu}U
             +U^{\dag}\nabla_{\mu}\nabla_{\nu}U.
\end{eqnarray*}
Then, from eq.(~\ref{e23}), we have
\begin{eqnarray}
\label{e26}
  ({\cal L}_{eff})_4&=&f_1<\nabla_{\mu}U^{\dag}\nabla_{\nu}U\nabla^{\mu}U^{\dag}
           \nabla^{\nu}U>+f_2<\nabla_{\mu}U^{\dag}\nabla^{\mu}U
           \nabla_{\nu}U^{\dag}\nabla^{\nu}U> \nonumber \\
          &\;&+f_3<\nabla_{\mu}\nabla_{\nu}U\nabla^{\mu}\nabla^{\nu}U^{\dag}>
           +f_4<\nabla_{\mu}\nabla_{\nu}U\nabla^{\nu}\nabla^{\mu}U^{\dag}>
           \nonumber \\
          &\;&+if_5<F^{\mu\nu}_L\nabla_{\mu}U\nabla_{\nu}U^{\dag}+
                 F^{\mu\nu}_R\nabla_{\mu}U^{\dag}\nabla_{\nu}U> \nonumber \\
          &\;&+f_6<F_{L\mu\nu}UF_R^{\mu\nu}U^{\dag}>+f_7<F_{R\mu\nu}F_R^{\mu\nu}
           +F_{L\mu\nu}F_L^{\mu\nu}>,
\end{eqnarray}
where $f_i(i=1,2,...,7)$ are coupling constants of $({\cal L}_{eff})_4$,
and they can be 
calculated explicitly from eq.(~\ref{e23})(~\ref{e18}) and (~\ref{e21})(see below).
At $O(p^4)$, EVRE, EVRV and EQCPM of Li model 
have been absorbed by $f_i (i=1,2...7)$.
  In order to transform eq.(~\ref{e26}) into the standard
lagrangian(~\ref{e7}), the following $SU(3)$ relation\cite{Gasser} should be used,
\begin{eqnarray}
\label{e27}
 &\;&<\nabla_{\mu}U^{\dag}\nabla_{\nu}U\nabla^{\mu}U^{\dag}\nabla^{\nu}U>
     \nonumber \\
 &=&-2<\nabla_{\mu}U^{\dag}\nabla^{\mu}U\nabla_{\nu}U^{\dag}\nabla^{\nu}U>
     +\frac{1}{2}<\nabla_{\mu}U^{\dag}\nabla^{\mu}U>^2  \nonumber \\
   &\;&+<\nabla_{\mu}U^{\dag}\nabla_{\nu}U><\nabla^{\mu}U^{\dag}\nabla^{\nu}U>.
\end{eqnarray}
In addition, using eq.(~\ref{e6}) we obtain
\begin{eqnarray}
\label{e28}
  &\;&<\nabla_{\mu}\nabla_{\nu}U\nabla^{\mu}\nabla^{\nu}U^{\dag}> \nonumber \\
  &=&<\nabla_{\mu}U^{\dag}\nabla^{\mu}U\nabla_{\nu}U^{\dag}\nabla^{\nu}U>
    +i<F^{\mu\nu}_L\nabla_{\mu}U\nabla_{\nu}U^{\dag}+
          F^{\mu\nu}_R\nabla_{\mu}U^{\dag}\nabla_{\nu}U> \nonumber \\
       &\;&-<F_{L\mu\nu}UF_R^{\mu\nu}U^{\dag}>+\frac{1}{2}<F_{R\mu\nu}F_R^{\mu\nu}
           +F_{L\mu\nu}F_L^{\mu\nu}>.
\end{eqnarray}
Inserting eq.(~\ref{e13})(~\ref{e26})(~\ref{e27}) into eq.(~\ref{e23})
we obtain the  desired  coupling constants $L_i$ and $H_1$ as follows,
\begin{eqnarray}
\label{e100}
  L_1&=&\frac{f_1}{2}  \nonumber \\
     &=&(\frac{1}{k}-\frac{F^2}{8})(\frac{1}{2}-\frac{2\gamma}{g^2})
       \frac{\alpha \beta^2}{4}+\frac{\gamma \beta^2}{8g^2}-\frac{ \beta^3}
       {32g}+\frac{ \beta^4}{128g^2} \nonumber \\
      &\;&+\frac{\gamma \beta}{2g}(1-\frac{ \beta}{g})^2
       (1-\frac{ \beta}{2g})+\frac{\gamma}{8}(1-\frac{ \beta}{g})^4, \\
\label{e101}
  L_2&=&2L_1,  \\
\label{e102}
   L_3&=&-2f_1+f_2+f_3+f_4                                    \nonumber \\
      &=&-(\frac{1}{k}-\frac{F^2}{8})(1-\frac{3\gamma}{g^2})
        \frac{3\alpha \beta^2}{4}
        -\frac{3\gamma \beta^2}{4g^2}+\frac{3 \beta^3}{16g^2}-
        \frac{3 \beta^4}{64g^2} \nonumber \\
       &\;&+\frac{\gamma}{2}(1-\frac{ \beta}{g})^2-4\frac{\gamma \beta}{g}
      (1-\frac{ \beta}{g})^2(1-\frac{ \beta}{2g})-\gamma(1-\frac{ \beta}{g})^4, \\
\label{e103}
   L_9&=&-f_5-f_3-f_4 \nonumber \\
      &=&-(\frac{1}{k}-\frac{F^2}{8})\frac{2\alpha\gamma \beta^2}{g^2}
       +\frac{\gamma \beta^2}{4g^2}+\gamma(1-\frac{ \beta}{g})^2
       +\frac{g \beta}{8}(1-\frac{\beta}{2g}), \\
\label{e104}
  L_{10}&=&f_6-f_3 \nonumber \\
      &=&(\frac{1}{k}-\frac{F^2}{8})(\frac{1}{4}-\frac{2\gamma}{g^2})
        \alpha \beta^2+\frac{\gamma \beta^2}{4g^2}-\frac{g \beta}{8}
        -\frac{\gamma}{2}(1-\frac{ \beta}{g})^2,  \\
  H_1&=&f_7+\frac{f_3}{2}=-\frac{L_{10}}{2}-\frac{g^2}{16}.
\end{eqnarray}

\begin{table}[hbp]
\centering
 \tabcolsep 0.2in
 \begin{tabular}{|c|c|c|c|c|c|}   \hline
    \hspace{0.4in}          &$ \chi PT[L_i^r(m_\rho)]$&
    \multicolumn{4}{|c|}{Li   Model}  \\ \cline{3-6}
         &  &$L_i^E$   &$L_i^V$&  $L_i^Q$&  Total       \\ \hline
   $L_1$   &$0.7\pm 0.3$ &$0.84$ &$-0.76$&$0.79$ &$0.87$ \\ \hline
   $L_2$   &$1.3\pm 0.7$ &$1.56$ &$-1.49$&$1.58$ &$1.75$ \\  \hline
   $L_3$   &$-4.4\pm 2.5$&$-3.54$&$2.59$ &$-3.17$&$-4.13$\\ \hline
   $L_9$   &$6.9\pm 0.7 $&$3.2$  &$-2.8$ &$6.3$  &$6.7 $ \\ \hline
   $L_{10}$&$-5.2\pm 0.3$&$-3.4$ &$1.57$ &$-3.17$&$-5.0$ \\ \hline
   $H_1 $    &             &       &       &       &$-5.2$ \\  \hline
   \end{tabular}
\begin{minipage}{5in}
\caption {\small
$L_i^r$ in units of $10^{-3}$, ${\mu}=m_\rho$.
 The $L_i^E$, $L_i^V$ and $L_i^Q$ denote effects of spin -1 meson resonances
 exchange,  effects of spin -1 meson  resonances vertices and effects of
 quark constituent
 of pesudoscalar meson contribute  to $L_i^r(i=1,2,3,9,10)$
 respectively.}
\end{minipage}
\end{table}

Above analytical expressions of $L_i (i=1, 2, 3, 9, 10)$ and $H_1$ are
main results of this paper.
  Remember that the Li model's inputs are $f_{\pi}=186MeV, g=0.35$ and
$m_{\rho}=770MeV$. Then due to eq.(~\ref{e19})(~\ref{e20})(~\ref{e22})
(~\ref{e25}), $F,\beta,\alpha,k$ are fixed
and $\gamma=\frac{N_c}{3(4\pi)^2},m_0^2=m_{\rho}^2g^2$.
The numerical results of $L_i(i=1,2,3,9,10)$ and $H_1$ are shown
in Table 1, which are in good agreement with their values in $\chi PT$
with renomalization scale parameter $\mu \sim m_{\rho}$. 
Thus, we have successfully derived $\chi$PT Lagrangian from Li model
without any additional assumptions.
This indicates that the low energy limit of Li model is 
equivalent to $\chi$PT.

In the rest of this section, we shall analyse contribution of
the EQCPM, EVRE and EVRV to chiral coupling constants $L_i(i=1,2,3,9,10)$
respectively. Consequently, $L_i$ of Li model are expressed as
$$ L_i=L_i^Q+L_i^E+L_i^V$$
where $L_i^Q, L_i^E$ and $L_i^V$ stand for contribution of EQCPM, EVRE and
EVRV respectively.
For this purpose, the lagrangian (~\ref{e14}) is separated into three parts,
$$ {\cal L}_{Re}={\cal L}_{Re}^{Q}+{\cal L}_{Re}^{E}
    +{\cal L}_{Re}^{V} $$
where
\begin{eqnarray}
  {\cal L}_{Re}^{E}&=&\frac{F^2}{16}<D_{\mu}UD^{\mu}U^{\dag}
              -\nabla_{\mu}U\nabla^{\mu}U^{\dag}>
              -\frac{g^2}{16}<2L_{\mu\nu}L^{\mu\nu}+2R_{\mu\nu}R^{\mu\nu}>
              \nonumber \\
              &\;&-\frac{3i}{\sqrt{2}}\gamma<\nabla_{\mu}U\nabla_{\nu}U^{\dag}
              L^{\mu\nu}+\nabla_{\mu}U^{\dag}\nabla_{\nu}UR^{\mu\nu}> \nonumber \\
              &\;&-\frac{g^2}{4\sqrt{2}}<L_{\mu\nu}F^{L\mu\nu}
              +2R_{\mu\nu}F^{R\mu\nu}>
              +\frac{1}{4}m_0^2<L_{\mu}L^{\mu}+R_{\mu}R^{\mu}>, \nonumber \\
  {\cal L}_{Re}^{Q}&=&\frac{F^2}{16}<\nabla_{\mu}U\nabla^{\mu}U^{\dag}>
           +\frac{\gamma}{2}<\nabla_{\mu}\nabla_{\nu}U\nabla^{\mu}
             \nabla^{\nu}U^{\dag}>    \nonumber \\
           &\;&-\frac{3i}{2}\gamma<\nabla_{\mu}U^{\dag}\nabla_{\nu}U
           F^{L\mu\nu}+\nabla_{\mu}U\nabla_{\nu}U^{\dag}F^{R\mu\nu}> \nonumber \\
           &\;&+\frac{\gamma}{4}<\nabla_{\mu}U\nabla_{\nu}U^{\dag}\nabla^{\mu}U
           \nabla^{\nu}U^{\dag}-2\nabla_{\mu}U\nabla^{\mu}U^{\dag}\nabla_{\nu}U
            \nabla^{\nu}U^{\dag}>, \nonumber  \\
  {\cal L}_{Re}^{V}&=&{\cal L}_{Re}-{\cal L}_{Re}^{Q}
         -{\cal L}_{Re}^{E},   \nonumber
\end{eqnarray}
There are no meson resonance fields in ${\cal L}_{Re}^Q$ which describe
the effects of quark constituent of pesudoscalar meson. But there are one or
two spin -1 meson resonances in ${\cal L}_{Re}^{E}$ which describe the
effects
of virtual resonances exchange(it is similar with Ref.\cite{Ecker}). The other
terms have been included in ${\cal L}_{Re}^{V}$ which describe the effects
of virtual resonances vertices. Three parts are all invariant under global
$U(3)_L \times U(3)_R$ chiral transformation consquently.
The EQCPM contributes to chiral coupling constants
$L_i^{Q}(i=1,2,3,9,10)$ and EVRE contribute to chiral coupling constants
$L_i^{E}(i=1,2,3,9,10)$ can be expressed as 
(see eq.(~\ref{e100})-(~\ref{e104})),
\begin{eqnarray}
   &&L_1^{Q}=\frac{\gamma}{8}; \;\;\;\;\;\;
   L_2^{Q}=\frac{\gamma}{4}; \;\;\;\;\;\;
   L_3^{Q}=-\frac{\gamma}{2};   \nonumber \\
   &&L_9^{Q}=\gamma; \;\;\;\;\;\;
   L_{10}^{Q}=-\frac{\gamma}{2}. \nonumber
\end{eqnarray}
and
\begin{eqnarray}
  L_1^{E}&=&(\frac{1}{k}-\frac{F^2}{8})(\frac{1}{2}-\frac{2\gamma}{g^2})
       \frac{\alpha \beta^2}{4}+\frac{\gamma \beta^2}{8g^2}
       +\frac{\gamma \beta}{2g}(1-\frac{ \beta}{2g}), \nonumber \\
  L_2^{E}&=&2L_1^{E},  \nonumber \\
  L_3^{E}&=&-(\frac{1}{k}-\frac{F^2}{8})(1-\frac{3\gamma}{g^2})
        \frac{3\alpha \beta^2}{4}-\frac{3\gamma \beta^2}{4g^2}+\frac{\gamma}{2}
       -4\frac{\gamma \beta}{g}(1-\frac{ \beta}{2g}), \nonumber \\
  L_9^{E}&=&-(\frac{1}{k}-\frac{F^2}{8})\frac{2\alpha\gamma \beta^2}{g^2}
       +\frac{\gamma \beta^2}{4g^2}
       +\frac{g \beta}{8}(1-\frac{\beta}{2g}), \nonumber \\
  L_{10}^{E}&=&(\frac{1}{k}
        -\frac{F^2}{8})(\frac{1}{4}-\frac{2\gamma}{g^2})\alpha \beta^2
        +\frac{\gamma \beta^2}{4g^2}-\frac{g \beta}{8} \nonumber
\end{eqnarray}
The numberical results of $L_i^{Q}$, $L_i^{E}$ and $L_i^{V}
(i=1,2,3,9,10)$ have been shown in table 1. As emphasision in Introduction
and section 2, both
the EQCPM and EVRV also contribute to $L_i(i=1,2,3,9,10)$ besides the EVRE
when spin -1 meson resonances were integrated out. Numberically, the
contribution of $L_i^Q+L_i^V$ to $L_i(i=1,2,3)$ are small, and $L_1^E,
L_2^E, L_3^E$ are dominant. But $L_9^Q+L_9^V$ and $L_{10}^Q+L_{10}^V$
are close to $L_9^E$ and $L_{10}$ respectively.
\vspace{8mm}

{{\bf 5. Discussion}}

It is necessary for meson theories in intermediate energy scale to examine
if its low energy limit is equivalent to $\chi$PT or not, because $\chi$PT
is the rigorous theory of QCD at low energies. In this paper, we have
done so for Li model, and proved that Li model is satisfied of this 
necessary condition. In the derivation of $L_i (i=1,2,3,9,10)$ and $H_1$,
all EVRE, EVRV and EQCPM at $O(p^4)$ are taken into account. The 
calculations presented in the paper are analytical and manifest without 
additional assumptions. They are beyond the similar calculations in 
refs.\cite{Ecker, Ecker1}.
The authors of ref.\cite{Ecker1} addressed that 
there are five conditions abstracted from QCD, which should be satisfied
for eligible chiral effective theories with meson resonances. Since
we have shown that the five constants $L_i (i=1,2,3,9,10)$ of Li model
are consistent with $\chi$PT, we conclude that Li model satisfies
these conditions. In refs.\cite{BAL1, BAL2,
BAL3, BAL4, Gao}, various aspects of the meson physics have been extensive
studied. Our investigations in this paper have proved further that
the dynamics of pseudoscalar, vector and axial-vector mesons in Li model 
is eligible. Consequently the successes of the phenomenology of
Li model become understandable more manifestly from QCD point of view.
Furthermore, there are some other models \cite{MB} in which
the $1^-$ and $1^+$ meson resonance fields are treated as ordinary vector
and axial-vector fields too(instead of as antisymmetric tensor fields like
in Ref\cite{Ecker}). Therefore the low energy limit of those models can
be derived and examined
by our method presented in this paper. The studies on them will be presented
elsewhere.

In the electromagnetic and weak interaction of mesons, the vector and
axial-vector mesons play essential role through VMD and AVMD. This 
was originally illustrated by Sakurai\cite{Sakurai}, and enjoys considerable 
phenomenological support. In the Li model, this is no longer an input, but 
a natural consequence of the model. Thus, our success in this paper 
indicates that VMD and AVMD are finely compatible with $\chi$PT.
 
Finally, we like to argue that in the Li model, besides the
exchange effects of virtual spin 1 meson resonances, the vertex effects
of them and the effects of quark constitutes of pseudoscalar mesons
contribute to $L_i (i=1, 2, 3, 9, 10)$ at $O(p^4)$ too. Therefore, 
in principle, the compatibility of VMD, AVMD with $\chi$PT in the chiral 
theories do not require that the values of $L_i (i=1, 2, 3, 9, 10)$
must be dominated by $V+A$ exchange effects only (like the case of
\cite{Ecker}).
In the view of the model presented in \cite{Ecker}, $L_i (i=1, 2, 3, 9, 10)$
are dominated by spin 1 meson resonance exchanges. The analysis
presented in this paper indicates that the result of \cite{Ecker} is
model dependent.

\vspace{1cm}

\begin{center} {\bf Acknowledgments} \end{center}

We would like to thank Dao-Neng Gao for helpful discussions, and
S.Randjbar-Daemi for discussions and carefully reading the manuscript.
MLY wishs to acknowledge the 
International Center for Theoretical Physics, Trieste,
for its hospitality where part of this work was done. This
work was supported in part by the National Funds of China through
C.N. Yang.

\end{document}